\documentclass[reprint,aps,prl,showpacs,preprintnumbers,amsmath,amssymb,amsfonts]{revtex4-1} 

\usepackage{graphicx}
 \usepackage{epstopdf}
 \usepackage{color}

\begin{document}

\title{Optical probing of shocks driven into overdense plasmas by laser hole-boring}

\author{N.~P.~Dover$^{1}$}
\author{C.~A.~J.~Palmer$^{1}$} 
\author{M.~ Babzien$^{2}$}
\author{A.~R.~Bell$^{3}$}
\author{A.~E.~Dangor$^{1}$}
\author{T.~Horbury$^{1}$}
\author{M.~Ispiriyan$^{4}$}
\author{M.~N.~Polyanskiy$^{2}$}
\author{J.~Schreiber$^{1,*}$}
\author{S.~Schwartz$^{1}$}
\author{P.~Shkolnikov$^{4}$}
\author{V.~Yakimenko$^{2}$}
\author{I.~Pogorelsky$^{2}$}
\author{Z.~Najmudin$^{1}$}

\affiliation{$^{1}$ Blackett Laboratory, Imperial College, London SW7 2AZ, United Kingdom}
\affiliation{$^2$ Accelerator Test Facility, Brookhaven National Laboratory, NY 11973, USA}
\affiliation{$^{3}$ Clarendon Laboratory, University of Oxford, Oxford OX1 3PU, United Kingdom}
\affiliation{$^{4}$ Stony Brook University, Stony Brook, NY 11794, USA}

\date{\today}

\begin{abstract}

Observations of the interaction of an intense $\lambda_{0} \approx 10\, \mu$m laser pulse with near-critical overdense plasmas ($n_e=1.8\,$-$\, 3\, n_c$) are presented. For the first time, transverse optical probing is used to show a recession of the front surface caused by radiation pressure driven hole-boring by the laser pulse with an initial velocity $>10^{6}\, \rm ms^{-1}$, and the resulting collisionless shocks.  The collisionless shock propagates through the plasma, dissipates into an ion-acoustic solitary wave, and eventually becomes collisional as it slows further. These conclusions are supported by PIC simulations which show that the initial evolution is dominated by collisionless mechanisms.
\end{abstract}

\maketitle
Collisionless shocks are associated with numerous natural phenomena from the bow-shock ahead of the earth's magnetosphere \cite{bow_shocks} to the blast wave in supernova explosions \cite{McKee:1974fv}. They can accelerate particles to high energies, and thus are implicated in the generation of cosmic rays \cite{cosmic_rays}. It has been suggested that collisionless shocks can be similarly used to create energetic particle beams in the laboratory. In particular, the radiation pressure of powerful lasers has been proposed as a shock-initiating piston \cite{DenavitPRL1992,shocks,macchi}. 
A number of measurements of energetic ion beams have been attributed to acceleration by the space charge fields created at the target front surface when an intense laser strikes a target with initial density greatly exceeding the critical density, $n \gg n_{c} \equiv \epsilon_{0} m_e \omega_{0}^{2} / e^{2}$, where $\omega_{0}$ is the laser angular frequency \cite{front_surf}.

In recent experiments, monoenergetic ion beams have been reported from targets of much lower (gaseous) density.  This was achieved by driving the target with longer wavelength ($\lambda_{0} \approx 10\, \mu$m) CO$_{2}$ lasers  \cite{PalmerPRL2011, HaberbergerNP2011}. Proton beams with energies from $\sim 1\,$ to 20 MeV were observed with energy spreads $< 5$\%. The acceleration was associated with shock-like structures moving away from the front surface of the plasma driven by the interaction with the laser.  Furthermore, in \cite{PalmerPRL2011} the most energetic beams were observed for $n$ approaching $n_{c}$. This is expected as momentum conservation implies the target surface's recession velocity should scale with mass density $\rho$ as $v_{hb}=\sqrt{(1+\eta)I/\rho c}$, where $\eta$ is the target reflectivity due to the radiation pressure of a laser of intensity $I$\cite{WilksPRL1992}.


The target front surface recession is called hole-boring and leads to a shock being driven into the target. Hole-boring has been experimentally identified from the Doppler shift of light back reflected from the interaction \cite{KodamaPRL1996} and its harmonics \cite{ZepfPoP1996}, and observed using x-ray refractometry \cite{TakahashiPRL2000}.  There have also been a number of studies of collisionless shocks generated by intense lasers, either directly through ponderomotive expulsion \cite{pond_exp} or from ablated plasma \cite{abl_shocks}, but these studies are generally limited to the underdense regime. Both x-ray studies and proton probing offer the ability to study to higher density but cannot directly measure density inside solid targets, where scattering is strong.

This letter details a study of the shock driven into a just-overdense plasma by laser hole-boring.
The plasma is driven by an infrared (CO$_2$) laser, for which $n_{c}\approx 10^{19} \rm \, cm^{-3}$; $> 100$ times lower than that for optical wavelengths. This allows direct \emph{optical} probing of the plasma. By varying the timing of the probe relative to the drive pulse, the shock evolution was studied, and three separate phases were identified. Initially the laser exerts a pressure on the opaque plasma, causing hole-boring and formation of either an electrostatic double layer at the surface, or a collisionless shock moving into the plasma, accelerating background ions.
Once the laser stops, particle reflection from the propagating shock structure causes it to decelerate rapidly, transitioning into a solitary ion-acoustic wave which continues to slow in the cooling plasma.  As the shock speed decreases, the ion mean free path also decreases and the shock eventually becomes collisional.
These findings are supported by particle-in-cell (PIC) simulations, which also help elucidate the mechanism of narrow energy spread ion beam generation.
  
The experiment was performed with the $\lambda_{0} \approx 10 \, \mu$m CO$_2$ laser at the Accelerator Test Facility at Brookhaven National Laboratory. The laser was circularly polarised. Streak images of the laser time profile showed a train of pulses, caused by spectral modulation in the amplifiers, each of pulse-length $\tau\approx 6$\,ps (FWHM) and 18\,ps pulse separation. On average 70$\%$ of the energy was in the leading two pulses, with a total integrated energy of 2\,-\,3\,J. The pulse was focused with an $f/3$ off-axis parabolic mirror to a spot size $w_0$ $\approx 70\,\mu$m (FWHM), giving a peak intensity $I \approx 5\times10^{15} \, \rm Wcm^{-2}$ and normalised vector potential $a_0$ $\approx$ 0.5. The laser was focused $\approx 0.7$\,mm above a 2\,mm diameter nozzle gas jet producing a helium gas neutral density profile, characterised by transverse interferometry \cite{NajmudinPoP2011}, approximated by an isosceles trapezoid of plateau width $1200\,\mu$m and side length $600 \, \mu$m.  

\begin{figure}  
 \includegraphics[width=0.47\textwidth]{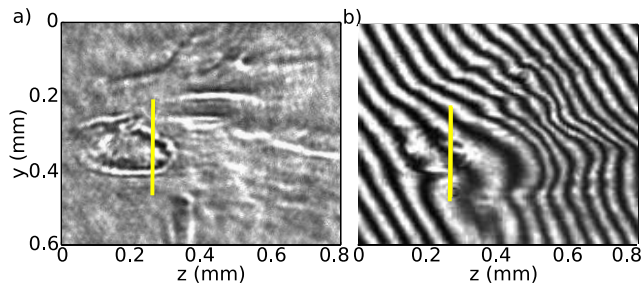}
 \caption{\label{fig1}a) Typical shadowgraphy and b) interferometry images for $t = 30$ ps, $n = 3\, n_{c}$. Vertical yellow lines indicate the initial critical density position. Laser enters from the left.}
\label{typical}
 \end{figure}

The probe, a frequency-doubled (532 nm) Nd:YAG laser beam of $\tau \simeq 6$ ps, was absolutely synchronised with the CO$_2$ pulse. The timing between them was varied using an optical delay line.  The probe passed orthogonally through the plasma before being split; with one part imaged directly for shadowgraphy, and the other going to a Mach-Zender interferometer, to diagnose plasma density. 
Examples are shown in Fig.~\ref{typical} for a peak initial density of $n_{e} = 3 \,n_{c}$, at $t=30\,$ps after the start of the interaction, which is just after the second major pulse in the pulse train has passed. Bright features in the shadowgram (Fig.~\ref{typical}a) highlight regions of rapidly changing refractive index and therefore density. The shadowgram shows a feature at the target front surface of $\sim 100 \, \mu$m transverse size  narrowing to $\sim 50\, \mu$m. Interferometry reveals that this is a high density feature with a region of considerably lower density inside it (fig.~1b),
which is the channel formed by the initial $75 \, \mu$m laser focal spot boring into the plasma.
The feature ends $\sim 45 \,  \mu$m longitudinally from the estimated initial critical surface, calculated from the gas density profile and supported by the off-axis plasma density profile.  Assuming the front originated at the critical surface implies an average $v = 1.5\times 10^{6} \, \rm ms^{-1} (\pm 30 \%) $, which agrees reasonably with $v_{hb} = 1.9\times 10^{6} \, \rm ms^{-1} $, obtained using the laser parameters for this shot and taking $\rho = m_{\rm He} n_{c}$.
This calculation ignores self focusing, the pulse train, and thermal effects.

 \begin{figure}[b]
 \includegraphics[width=0.47\textwidth]{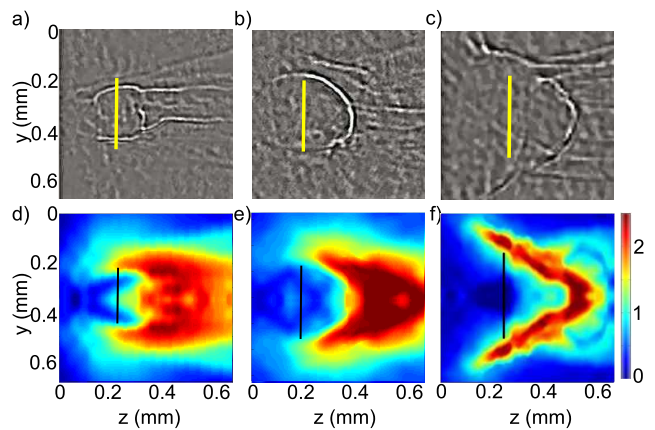}
 \caption{\label{fig2} (colour online) Time series of shadowgraphy (top) with corresponding density maps (below) for a,d) t = 200, b,e) 500 and c,f) 1600\,ps.  The density is in units $n_c$. Laser enters from the left. {The shadowgraphy images have been Fourier filtered for clarity.  The initial critical surface is indicated with a vertical line for all images.}  For all shots $n_{e} = 1.8 \, n_c$.}
 \end{figure}
 
Fig.~\ref{fig2} shows a series of shadowgrams taken with varying probe timings up to $t=1600 \,$ps. The front seen in Fig.~\ref{typical}a remains clearly defined, moving further into the plasma as time progresses. Horizontal features can be seen further into the plasma, particularly for $t=200 \,\rm ps$ (fig.~\ref{fig2}a). These are likely to have been created as the target was being ionised, since the gas target is initially transparent to the ionising laser.
Below the shadowgraphy images are the corresponding density maps derived from interferometry (fig.~\ref{fig2}d-f).   The phase shift was assumed to be symmetrical around the laser axis, allowing Abel inversion to retrieve the density profile. 
The accuracy of the recovered density maps is limited by asymmetry in the interferograms, caused by, for example, slight vertical density gradients and laser non-uniformities. 

\begin{figure}  
 \includegraphics[width=0.47\textwidth]{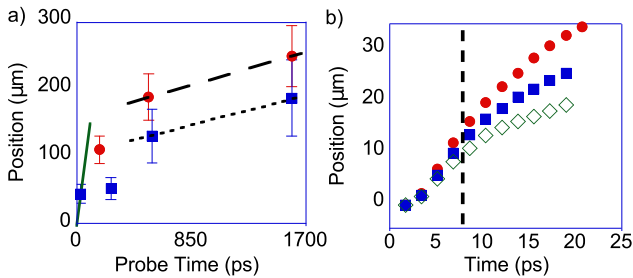}
 \caption{\label{fig3} (colour online) a) Shock propagation distance $z_{sh}$ with time from shadowgrams for $n_e \in (1.8, 3) \, n_c$ (red circles, blue squares).  Green solid line indicates initial $v_{hb}$, and black (dashed, dotted) lines represent the mean final $v_{sh}$ for $n_{e} \in (1.8, 3)\, n_c$; b) $z_{sh}$ from 2DPIC for $n \in (1.8, 3, 6) \, n_c$ (red circles, blue squares, green diamonds).  The laser stops at $t \approx 7.5$ ps, indicated by vertical dotted line}
 \end{figure}

Once sufficiently ionised, the plasma becomes opaque ($n_{e}>n_{c}$) and the laser is reflected at the critical surface.  It then exerts radiation pressure on the plasma, launching a shock into the initially relatively cold plasma.  The peak density associated with this moving front exceeds the initial background plasma density (fig.~\ref{fig2} b,c), and shows a steepening in the feature profile even at late times; clear indication of shock formation. Two probe timing scans were taken for peak densities $n_{e} \in (1.8, 3)\, n_c$.  The distance of shock propagation from the initial critical surface position thus obtained is plotted in fig.~\ref{fig3}a as a function of time.   The shock initially moves rapidly into the plasma before slowing at later times, with the shock velocity initially higher for lower density. The mean velocities for different parts of the shock trajectory are indicated on the plot. The previously estimated hole-boring velocity ($v_{hb} \approx1.5 \times 10^{6} \, \rm ms^{-1}$) is indicated, slowing to $v_{sh} \approx  5 \times 10^{4} \, \rm ms^{-1}$ for $t=500\,$-$\,1700 \,$ps, for both $n_{e} \simeq 1.8\,$  and $3 \, n_{c}$.  The ion mean-free-path due to (ion-ion) collisions for ions with $v=v_{hb}$ is $\ell_{mf}\approx 30 \,$mm  \cite{Spitzer1962}.  This is much larger than the scale of the measured features  ($<100\, \mu$m) and even the gas jet, indicating that the shock must initially be collisionless.  At the latest times measured, as the shock slows, $\ell_{mf} (\propto v^{4})$ decreases significantly, and collisions play a role in dissipation. 
  
To further investigate this behaviour, simulations of the interaction were performed using particle-in-cell (PIC) codes. 2D3V simulations were performed with the code {\sc osiris}  \cite{Fonseca2002}.  A circularly polarised $\lambda = 10.6 \, \mu$m pulse, with $\tau = 6$ ps, $a_0=0.55$, $w_L=$ 70 $\mu$m was focused on a fully ionised, initially cold ($T_e=T_i=0$) He plasma of density $n_{e} = 1.8\,$-$\,6 \, n_c$ with a step density gradient.  

\begin{figure}[b]
 \includegraphics[width=0.47\textwidth]{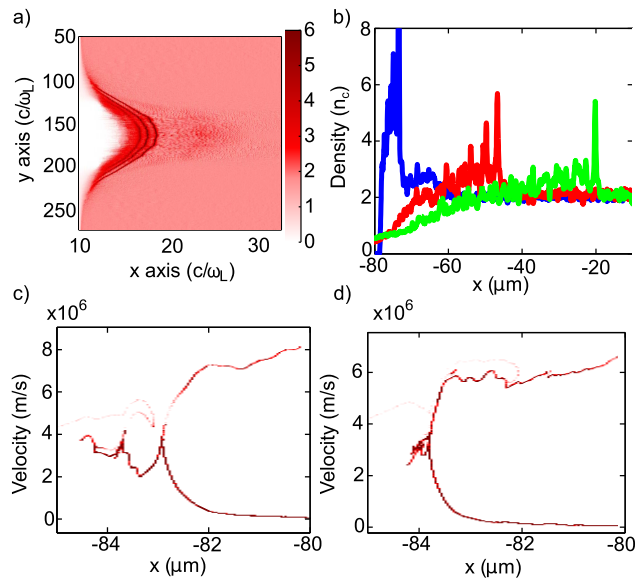}
 \caption{\label{fig4} (colour online) a) Density distribution from 2DPIC for $n_{e} = 1.8n_{cr}$, $T_{ei}=0$ at $t = 9$ ps.  b) Evolution of 1D He$^{2+}$ charge density for $n_{e} = 2\,n_{cr}$, $T_{ei} = 5\,$keV,  at $t=5$ ps (blue, corresponds to end of laser pulse), 15 ps (red) and 30 ps (green). $v_{x} x$ phase space for He$^{2+}$ from 1DPIC at the peak of laser for c) $T_{ei} = 5$ and d) $T_{ei} = 1$ keV.      }
 \end{figure}

Fig.~\ref{fig4}a shows the plasma density profile from the simulation just after the laser pulse. It shows a density increase at the front along with the cavity formed by the hole-boring, in agreement with what is seen experimentally.
The front is formed by radiation pressure displacing plasma electrons to a collisionless skin-depth.
The ions, which due to their greater mass are unaffected by the laser fields, accumulate at the front due to the space-charge field of the ensuing electrostatic double layer \cite{BlockAaSS1978}. 
The large electrostatic field associated with this front also accelerates upstream ions. This leads to the formation of a large population of energetic ions that can be seen to outrun the shock front in fig.~\ref{fig4}a.

Since the plasma ahead of the front is initially cold, hole-boring causes a compression of the plasma.
The piston-like effect of the radiation pressure launches a shock into the plasma. 
The motion of the shock front during the simulation is shown in fig.~\ref{fig3}b for three different starting $n_{e}$. 
Although initialised cold, the plasma is heated by vacuum heating \cite{BrunelPRL1987}, due to the buckling of the front surface from hole-boring.  Hence for lower $n_{e}$, where hole-boring is faster, the rate of heating increases more rapidly, and the shock moves faster.

To obtain high resolution, the simulation box for 2D simulations was limited to $60\, \mu$m longitudinally. As a result a sheath quickly forms at the rear of the simulation box due to high energy electrons that leave the box.  This sheath reflects electrons back into the plasma much more efficiently than in the experiment, where the plasma size was $> 500 \, \mu$m. This results in an unphysically high background temperature $T_{e}$, and consequently poor modelling of the shock front behaviour.
To model the shock evolution with a more realistic box size, a series of 1D PIC simulations were performed with the code {\sc epoch1d}.  A fully ionised He plasma with a top hat profile of $n_{e} = 2  n_c$ was irradiated with an $a_0 = 1$ circularly polarised laser, $\tau_{L} = 5$ ps (gaussian) pulse with $\lambda=10.6 \, \mu$m. The simulation box was $200\, \mu$m, with 5000 cells and $2.5\times10^{5}$ particles per species.  The ions were set to be initially cold ($T_{i} = 0$ eV). 1D simulations cannot model the buckling of the plasma due to hole-boring. Therefore, to assess the influence of heating, the initial electron temperature was varied between $T_{ei} = 0\,$-$\,50$ keV.
These simulations are comparable to those in \cite{DenavitPRL1992}, as the lower intensity here is compensated by reduced densities to give similar piston velocities, and also in \cite{macchi}, which discusses the role of ion temperature on the shock structure.


Whilst the laser is on, the laser bores a hole into the plasma irrespective of $T_{ei}$, although the hole-boring speed decreases slightly for higher temperatures due to the increase in the electron thermal pressure.  For the highest $T_{ei}$ (= 50 keV), where $v_{hb}$ is only just higher than the initial sound speed, $c_s = \sqrt{kT_e/m_i}$, a series of solitary waves are pushed into the plasma.  Decreasing the initial temperature of the simulation to 5 keV increases the ratio of piston velocity to the local sound speed, and the ion density profiles (e.g.~fig.~\ref{fig4}b) show the formation of a collisionless electrostatic shock. The shock moves at $v_{sh} \approx 4 \times 10^6 \, \rm ms^{-1}$, compared with $v_{hb} \approx 3  \times 10^6 \, \rm ms^{-1}$, and so moves ahead of the critical surface. The density at this shock can be as high as $n_{e} = 10\, n_{c}$.  
Fig.~\ref{fig4}b shows that the shock width ($\sim c/\omega_{p}$) is less than 1 $\mu$m, which is why though it is clearly visible from shadowgraphy (\ref{fig2} a\,-\,c), it is not properly resolved in the interferograms (\ref{fig2} d\,-\,f).
The potential of this front ($0.15 \, \rm MV$) is so large that the associated electric field, which reaches $> 200 \, \rm GVm^{-1}$, efficiently reflects nearly all upstream ions. Phase space diagrams (fig.~\ref{fig4}c) show that the reflected ions gain a speed of $2v_{sh}$ by `bouncing-off' the shock front. Whilst the laser is active, it is $v_{sh}$ that determines the accelerated ion velocity (not $v_{hb}$). This results in a remarkably narrow energy spread for the reflected ions \cite{PalmerPRL2011}.  With $T_{ei} = 5\,$keV, there is partial reflection and partial transmission through the shock front. The transmitted ions display the characteristic for a shock solution with partial particle reflection, displaying numerous trailing perturbations in the potential, as is evident in Fig.~\ref{fig4}c (and \ref{fig4}a) \cite{shock_sims}. 
Further reducing $T_{ei}< 1 \,$keV (so $v_{hb} \gg c_s$, fig.~\ref{fig4}d), a region of high ion density is formed right at the hole-boring front, creating a double layer moving at $v_{sh} \approx v_{hb}$,  which reflects nearly 100$\%$ of upstream ions.  

Therefore, the simulations show three separate interaction regimes for $T_e \gg T_i$: 1) for $v_{hb} \lesssim 3 c_s$, the laser launches solitary waves into the plasma from which there is no particle reflection; 2) for $v_{hb} \gtrsim 3 c_s$, a shock forms moving ahead of the hole-boring front with partial particle reflection; and 3) for $v_{hb} \gg c_s$  there is 100$\%$ reflection and $v_{sh} \sim v_{hb}$ \cite{shock_sims,DenavitPRL1992}.  When considering the interaction as a laser plasma ion source, there is therefore a tradeoff between higher maximum energies from a hotter plasma, where a small fraction of the particles are accelerated up to higher energies, and higher efficiency from a colder plasma where all the particles are accelerated to a smaller maximum energy. This explains the larger flux of accelerated protons observed with a short, circularly-polarised laser beam \cite{PalmerPRL2011}, as opposed to a longer, linearly-polarised driver where heating is stronger \cite{HaberbergerNP2011}.

At the end of the laser pulse, the potential associated with the shock is still sufficient to reflect ions.  However, as there is no longer a pressure being exerted by the laser, this particle reflection reduces the energy of the ion front, causing it to slow until reaching a critical Mach number, $M_{s}$, where particle reflection practically stops. Assuming $T_i = 0$, one expects a lower limit of $M_s = 1.6$ given an isothermal approximation for the plasma, whereas assuming maximum electron trapping behind the shock gives an upper value of $M_s \sim 3$ \cite{shock_sims}. In the simulations, at the end of the laser pulse the shock speed drops to $v_{sh} \sim 3 M_{s} c_{s}$. For the `cold' simulations ($T_{ei} = 0$ keV), this results in the shock practically stopping, not matching our observations. 

 

For simulations with a $T_{ei} \neq 0$, $T_{e}$ decreases with time as energetic electrons leave the simulation box, causing the soliton speed to exceed $M_{s} c_{s}$. This stimulates further particle reflection until $v_{sh} \sim M_{s} c_{s}$ again. Though artificial, this does partly imitate reality, since the shock will lose energy as it expands transversely whilst moving forward.
$M_{s}$ is seen to decrease from 3 to $<2$ during the simulated shock propagation, suggesting a transition between the maximal trapping limit and the isothermal limit.  
By the end of the simulation ($t=95\,$ps), the ion soliton is still visible moving in the plasma with a coherent structure.  Therefore, in a cooling plasma $v_{sh}$ constantly decreases to maintain $M=M_{s}$, which is itself decreasing.  Eventually $v_{sh}$ reduces such that $\ell_{mf}$ becomes less than the shock thickness, and the shock becomes collisional. Steepening of the shock profile is observed again at very late times due to collisions (fig.\ref{fig2}). This phase is not modelled in the PIC simulations.

In conclusion, the dynamics of the shock driven into an overdense target due to hole-boring have been observed, from the double-layer to ion acoustic soliton and finally to collisional phases, for the first time experimentally. These results hence provide insight into the dynamics of shock acceleration in plasmas, which may prove to be an interesting route to high energy ion production.

$^{*}$ Present address: Fakult\"at f\"ur Physik, Ludwig-Maximilians-Universit\"at M\"unchen and Max-Planck-Institut f\"ur Quantenoptik, Garching, Germany

\begin{acknowledgements}
The work was part funded by RCUK Libra Basic Tech and US DOE DE-FG02-07ER41488 grants.  We thank the {\sc Osiris} consortium (UCLA/IST) for use of {\sc Osiris} and K. Kusche and the ATF technical staff for experimental assistance.
\end{acknowledgements}

\end{document}